\documentclass[sigconf,natbib=true, nonacm=true]{acmart}

\usepackage{subcaption}
\usepackage{booktabs}
\usepackage{multirow}

\definecolor{diagram_green}{RGB}{85, 147, 132}
\definecolor{diagram_blue}{RGB}{46, 95, 127}
\definecolor{diagram_orange}{RGB}{239, 137, 51}

\AtBeginDocument{%
  }

\begin{document}

\title{Reverse-Engineering the Retrieval Process in GenIR Models}

\author{Anja Reusch}
\email{anja@campus.technion.ac.il}
\orcid{0000-0002-2537-9841}
\affiliation{%
  \institution{Technion – Israel Institute of Technology}
  \city{Haifa}
  \country{Israel}
}

\author{Yonatan Belinkov}
\email{belinkov@technion.ac.il}
\affiliation{%
  \institution{Technion – Israel Institute of Technology}
  \city{Haifa}
  \country{Israel}
}

\begin{abstract}
Generative Information Retrieval (GenIR) is a novel paradigm in which a transformer encoder-decoder model predicts document rankings based on a query in an end-to-end fashion. These GenIR models have received significant attention due to their simple retrieval architecture while maintaining high retrieval effectiveness.
However, in contrast to established retrieval architectures like cross-encoders or bi-encoders, their internal computations remain largely unknown. Therefore, this work studies the internal retrieval process of GenIR models by applying methods based on mechanistic interpretability, such as patching and vocabulary projections.
By replacing the GenIR encoder with one trained on fewer documents, we demonstrate that the decoder is the primary component responsible for successful retrieval. Our patching experiments reveal that not all components in the decoder are crucial for the retrieval process. More specifically, we find that a pass through the decoder can be divided into three stages:
(I) the priming stage, which contributes important information for activating subsequent components in later layers; (II) the bridging stage, where cross-attention is primarily active to transfer query information from the encoder to the decoder; and (III) the interaction stage, where predominantly MLPs are active to predict the document identifier. Our findings indicate that interaction between query and document information occurs only in the last stage.
We hope our results promote a better understanding of GenIR models and foster future research to overcome the current challenges associated with these models.\footnote{Code and models can be found here: \url{https://github.com/technion-cs-nlp/re-gen-ir}.}
\end{abstract}

\maketitle

\section{Introduction}

Generative Information Retrieval (GenIR)~\cite{metzler2021rethinking} trains a neural model to associate a query with a document identifier that satisfies the information need expressed within the query~\cite{tay2022transformerdsi}. This approach applies transformer models (usually an encoder-decoder model) as an end-to-end retrieval system.
The introduction of this new retrieval paradigm has received much attention~\cite{pradeep2023does, wang2022neuralnci, bevilacqua2022autoregressiveseal, nguyen2023generative, zhou2022ultron, liu2024robustness}, due to the simplified architecture and the fact that only one pass (or a few passes) through the model is required. 
Despite these advantages, GenIR models are commonly applied as a black box, obscuring the process by which they perform retrieval. %

\begin{figure}
    \centering
    \includegraphics[width=0.99\linewidth]{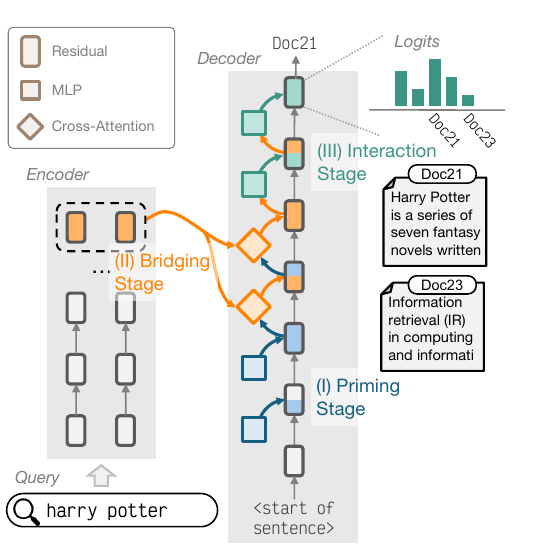}
    \caption{A simplified view of the retrieval process in the Generative IR models in this work. After the encoder processes the query, the decoder operates in three stages: (I) the \textcolor{diagram_blue}{Priming Stage}, where MLPs activate to prepare the residual stream triggering the cross-attention components in (II) the \textcolor{diagram_orange}{Bridging Stage}, which transfer query information from the encoder to the decoder's residual stream; and (III) the \textcolor{diagram_green}{Interaction Stage}, in which MLPs process the query information from the cross-attention to adjust the logits, promoting relevant documents.}
    \label{fig:overview-figure}
\end{figure}

In contrast,  past research on cross-encoders~\cite{chen2024axiomatic,zhan2020analysis} and dense retrieval models~\cite{wallat2024causal,ram2023you,Formal2021WhiteBox} has shed light on how these models judge the relevance of a document given a query. 
Another line of research focuses on the inner workings of transformer decoder models (LLMs) which are recently popularized due to their success in text generation. Here, research has identified key components that are crucial for performing certain tasks (e.g., \cite{elhage2021mathematical,wang2023interpretability,hanna2024does,hou2023towards}) and used these insights to improve model performance~\cite{meng2022locating,merullocircuit}. 
Despite these efforts, their insights are not applicable to GenIR models due to differences in training objective and model architecture. %

Therefore, this work studies the retrieval process within GenIR models. We investigate the following questions: Which components are responsible for predicting the correct document identifier and how do they interact with each other?
First, we show that the GenIR encoder can be replaced with encoders that were not trained on the target documents while still leading to successful retrieval by the trained decoder (Sec.~\ref{sec:encoder-swap}). This finding indicates that the encoder does not need to be specialized for the retrieval dataset or task, suggesting that retrieval happens primarily in the GenIR decoder.  
Then, we analyze the pass through the decoder with three types of methods (Sec.~\ref{sec:components}): (a) causal interventions, where we modify activations of different components  %
to identify which components are essential for retrieval; (b) vocabulary projections, where we project activations to the space of the vocabulary, which includes document identifies, to measure which components directly contribute to the ranking of the document identifier; and (c) interaction studies, where we assess how different components interact to perform the final ranking.

Our findings are summarized in Fig.~\ref{fig:overview-figure}. We discover that the pass through the decoder can be grouped into three stages: (I) the priming stage, which is query-agnostic, but contributes important information for the activation of components in later layers; (II) the bridging stage, in which cross-attention components---triggered by the components in Stage I---transfer query information from the encoder to the decoder; and (III) the interaction stage, which is the first time that document-specific information interacts with query information. We demonstrate that this flow exists in several GenIR models trained on Natural Questions~\cite{kwiatkowski2019natural} and TriviaQA~\cite{JoshiTriviaQA2017}.

The main contributions of this work are summarized as follows:
\begin{itemize}
    \item Reverse-engineering the retrieval process in GenIR models,
    \item Identifying the role and interaction of MLPs and Cross Attention within the retrieval process, and
    \item Release of our code and trained models for further evaluation and analyses.
\end{itemize}

\section{Background}
\label{sec:background}
\subsection{Insights into Language Models}

\begin{figure}[t]
    \centering
    \includegraphics[width=0.99\linewidth]{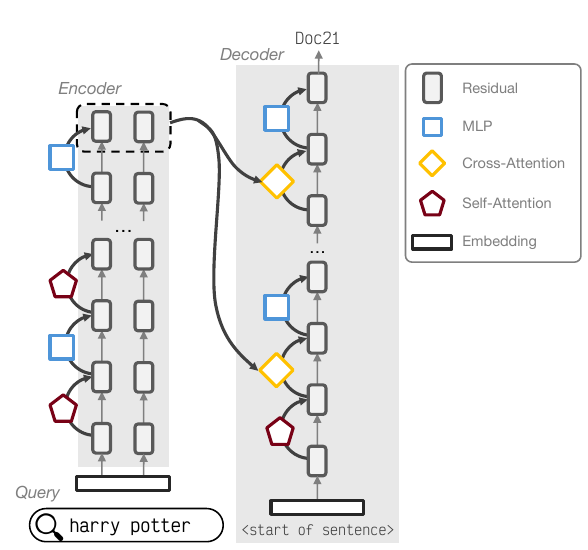}
    \caption{Overview of a transformer encoder-decoder for GenIR, depicted as the residual stream to which components read and write.}
    \label{fig:transformer-overview}
\end{figure}

A large body of work has attempted to uncover the internal mechanisms of Transformer models~\cite{elhage2021mathematical, hanna2024does, hou2023towards, vig2020investigating, meng2022locating,wang2023interpretability, chen2024axiomatic}. This field, known as mechanistic interpretability, is concerned with reverse engineering how language models implement different functionalities. A prominent view in this field is the residual stream view as introduced by Elhage et al.~\cite{elhage2021mathematical}. Here, the pass through the transformer model can be seen as sequence of operations that read from and write to a stream of vectors called the residual stream. The idea is that model components such as attention or multilayer perceptrons (MLPs) get as their input the residual stream and on the contrary their output is added to the residual stream. Each model component contributes to this residual stream in this view. Fig.~\ref{fig:transformer-overview} visualizes this idea in the context of GenIR. 

Subsequent work identified sub-graphs of a model's computation graph, called circuits, which are responsible for solving a certain task, e.g., indirect object identification~\cite{wang2023interpretability, merullocircuit},  mathematical~\cite{hanna2024does} or reasoning tasks~\cite{hou2023towards}.
To locate relevant model components, many of these studies apply mechanistic interpretability methods referred to as ``activation patching'' or ``causal mediation analysis'' \cite{vig2020investigating, meng2022locating, geiger2021causal,wang2023interpretability}. 
The core idea is to replace the activations of model components with different activations and measure the impact of this replacement. For example, to analyze the prompt ``The Eiffel Tower is in~\_\_'', we replace the activation of each model component one by one with the activations for the prompt ``The Colosseum is in~\_\_''. If the model was still able to predict ``Paris'' after a component's activation was replaced, this component was not responsible for predicting the output token. If the prediction switched to ``Rome'', the component was crucial for the model's computation of this particular output. By doing so, it is possible to identify all components crucial for a certain task, and reverse-engineer the circuit responsible for the models behavior.

However, the majority of this recent work analyzes decoder-only models and does not study retrieval scenarios. Since the prevailing approaches in information retrieval focus on encoder-only or encoder-decoder models, which are trained for different objectives, techniques and insights from these works might not be directly applicable. A few recent studies started working on language models in the context of retrieval: \cite{wallat2024causal} analyze how dense retrievers apply common retrieval properties, \cite{chen2024axiomatic} apply activation patching in a cross-encoder scenario, \cite{ram2023you} show that dense retrievers implement query expansion, \cite{Formal2021WhiteBox} performed an analysis of ColBERT~\cite{khattab2020colbert}, and \cite{zhan2020analysis} analyzed the interaction of query and document in a cross-encoder based on BERT~\cite{devlin2019bert}.
While these studies led to profound insights into the retrieval behavior of transformer models in dense retrieval or cross-encoder scenarios, none of them dealt with GenIR or even encoder-decoder models. Therefore, to the best of our knowledge, this is the first work to study GenIR models mechanistically.

\subsection{Generative Information Retrieval}

Generative Information Retrieval (GenIR) refers to the relatively new paradigm of using a transformer model in an end-to-end fashion to rank documents given a query. The model, usually an encoder-decoder,\footnote{A recent work~\cite{tang2024self} used Llama~2~\cite{touvron2023llama}, a decoder model. In this scenario the model generates the entire passage, which is closer to the text generation objective of decoder-only models compared to other GenIR approaches where document identifiers are generated.} is trained to associate the content of documents with their document identifiers. This approach was popularized by Tay et al.~\cite{tay2022transformerdsi}, who provided a proof-of-concept that their approach, DSI, is able to outperform comparable dense retrieval models. 
DSI fine-tuned T5 encoder-decoder models~\cite{raffel2020exploring} in a multi-task fashion: They applied an indexing phase where the model predicts the associated document identifier given the document input, and a retrieval phase where the model gets a query as its input and outputs the document identifier. Document identifiers were represented as either a new atomic token or as a sequence of tokens. As we focus on atomic document identifiers, we not cover other versions and refer the reader to~\cite{tay2022transformerdsi} for further details on them.

Fig.~\ref{fig:transformer-overview} depicts a GenIR  model. The input to the encoder is a query of $N$ token. The encoder itself employs the same architecture as popular encoder-only models such as BERT and outputs embedding vectors $e_1, \ldots, e_N$. The decoder contains cross-attention components in each layer in addition to attention and MLPs, which are also found in decoder-only models. The input to the decoder is a [start-of-sentence] token. In each layer, three components add information to the residual stream: attention, cross-attention, and MLPs~\cite{vaswani2017attention}. %
Each MLP component consists of two feed-forward layers FF$^\text{proj}$, and FF$^\text{out}$ connected by a ReLU non-linearity:
\begin{equation}
\label{eq-MLP}
\text{MLP}(r) = \sum_i \text{FF}^\text{out}_i \cdot a(r)_i, \text{ with } a(r)_i = \text{ReLU}(\text{FF}^\text{proj}_i \cdot r), 
\end{equation}
where $i$ denotes the $i$th neuron, $a(\cdot)_i$ its activation, and FF$_i$ the $i$th column/row of the linear layer weight matrices. T5 does not use bias terms in feed-forward layers. Eq.~\ref{eq-MLP} allows us to view the activation as a dot product of the input to the MLP, i.e., residual stream hidden state $r$, and each neuron's row in FF$^\text{proj}$, which gets passed through the ReLU activation. Since the ReLU replaces negative values by 0, only positively activated neurons contribute to the MLP output. 

Each cross-attention head has two inputs: the residual stream $r$ and the output of the encoder $e_1, \ldots, e_N$. 
These vectors get multiplied by the key matrix $W_K$ forming the key vectors $k_1, \ldots, k_N$. Similarly, the residual stream $r$ is multiplied by the query matrix $W_Q$ forming the query vector $q$:
$$W_K \cdot (e_1, \ldots, e_N) = (k_1, \ldots, k_N); \quad W_Q \cdot r = q$$
The dot product of query and keys results in similarity scores $s_i$ for each key $k_i$ with regard to the query $q$: $$(s_1, ..., s_N) = q^T \cdot (k_1, ..., k_N)$$ In the calculation of the cross-attention head, these similarity scores are transformed by a softmax with additional scaling:
$$(a_1, \ldots, a_N) = \text{scaled-softmax}(s_1, \ldots, s_N)$$
These cross-attention scores $a_i$ serve as weights in the output $o$: $o = \sum_i^N a_i \cdot v_i$, where $v_i = W_V \cdot e_i$ is a value vector. Each cross-attention component consists of $H$ heads, which each have separate $W_Q$, $W_K$, and $W_V$, resulting in one output vector $o_h$ per head $h$. These output vectors are concatenated and multiplied by the output matrix $W_O$ resulting in the cross-attention output: $$
\text{Cross-Attention}(r) = [o_1, \ldots, o_H] \cdot W_O$$

The (self-)attention component works in a similar way, but instead of using the encoder output as key and value vectors, it uses the decoder residual steam as keys and values. Since the residual stream consists of only one vector (because in the case of \textit{atomic} GenIR models, the decoder receives only one token as its input), self-attention only transforms the residual stream vector linearly and adds it back to the residual stream.

After the pass through all layers, the residual stream is multiplied by an ``unembedding matrix'' $W_U \in \mathrm{R}^{d_V \times d_\text{model}}$, which maps the residual stream $r \in \mathrm{R}^{d_\text{model}}$ to the dimensionality of the vocabulary $d_V$: $W_U \cdot \sqrt{d_{model}} \cdot \text{ReLU}(r) = l$. The output $l \in \mathrm{R}^{d_V}$ are the logits for each token in the vocabulary. Since we adopt the atomic document identifier approach from DSI, we receive one logit for each document identifier along with the logits of the encoder-decoder model's word vocabulary (henceforth: ``non-document-identifiers''). These logits can be converted to a probability distribution over the vocabulary by using the softmax. %

Following the success of DSI, other GenIR versions emerged~\cite{zhuang2022bridging, wang2022neuralnci, bevilacqua2022autoregressiveseal, nguyen2023generative, zhou2022ultron}.
Most of these adaptions aim to solve the underlying issue of GenIR models: the limited corpus size. An extensive comparison of GenIR variants~\cite{pradeep2023does} found that GenIR models perform worse compared to traditional methods when scaled to corpora of over eight million documents. Recently, also questions on the robustness of GenIR models~\cite{liu2024robustness} and the ranking quality~\cite{chen2023understanding,zhou2024roger} arose. 
Establishing a better understanding of the inner workings of GenIR models might help in solving these issues. 

\section{Training and Analysis Setup}
\label{sec:setup}
Since prior work did not publish trained GenIR models, we start by training models based on the DSI setup~\cite{tay2022transformerdsi}. Our models use atomic IDs, direct indexing with a maximum token length of 32, and an indexing-to-retrieval ratio of 32. We fine-tune T5-large models on Natural Questions~\cite{kwiatkowski2019natural} in three different scenarios, with 10k and 100k unique documents (queries on 1k documents for each validation and test set, the rest for training), and the entire Natural Questions dataset denoted by NQ320k. %
To verify our findings in another scenario, we additionally train two T5-large models on Trivia-QA~\cite{JoshiTriviaQA2017}, one using the same scenario as DSI (Trivia-QA) and one without a dedicated indexing stage, but instead using generated queries provided by~\cite{wang2022neuralnci} (Trivia-QA QG). We use the data splits provided by the data set. We set all hyperparameters according to the ones reported in~\cite{tay2022transformerdsi} and train using Huggingface Transformers~\cite{wolf-etal-2020-transformers}. Patching experiments as well as logit analyses are performed using TransformerLens~\cite{nanda2022transformerlens}.
The models' performance on our test sets can be viewed in Tab.~\ref{tab:model-results}. Our results are in line with those reported by previous work~\cite{tay2022transformerdsi,pradeep2023does}.

\begin{table}[t!]
\caption{Results on the respective test set for the models trained in this work. %
}
\label{tab:model-results}
\begin{tabular}{@{}lrrr@{}}
\toprule
 Dataset      & Hits@1 & R@5 & Hits@10 \\ \midrule
NQ10k        & 40.4  &  58.1   &  63.9    \\
NQ100k       & 22.6   & 42.8 &   50.2  \\
NQ320k       &   20.0     &  39.8   &  48.0       \\
Trivia-QA    &   60.4     &  64.2   &   82.2      \\
Trivia-QA QG &   49.8     &   51.7  &  67.0  \\\bottomrule
\end{tabular}
\end{table}

\section{The Role of the Encoder in GenIR}
\label{sec:encoder-swap}

Generative IR models usually employ a transformer encoder-decoder architecture. The decoder is responsible for generating the output token depending on what was processed by the encoder. Therefore, the decoder needs to store some information about the document tokens in order to output the correct token. However, what is the role of the encoder? Is information about the corpus on which the model was trained encoded in the weights of the encoder?

To investigate this, we design an experiment where the encoder is trained on fewer documents than the decoder and test if retrieval can still be performed using this ``incomplete'' encoder.
In total, we train four additional GenIR models on NQ10k, but omit ten different documents (taken from the validation set) when training each model. 
Then, we use the ``full'' GenIR model, which was trained on all NQ10k documents (including the $4 \times 10$ documents), but replace its encoder with an encoder of a model that was trained with ten missing documents. We feed the query of the removed documents to this ``hybrid'' encoder-decoder to measure how well the correct document can still be predicted. If information on the documents is stored exclusively in the encoder, the hybrid model, with an encoder that was not trained on these ten particular documents, should fail to correctly retrieve the removed documents. 
Before swapping the encoder, we measure the performance of the four models (with their own encoder and decoder). They obtain comparable performance to the ``full'' model, which was trained on all documents.

\begin{figure*}
    \centering
    \includegraphics[width=0.99\linewidth]{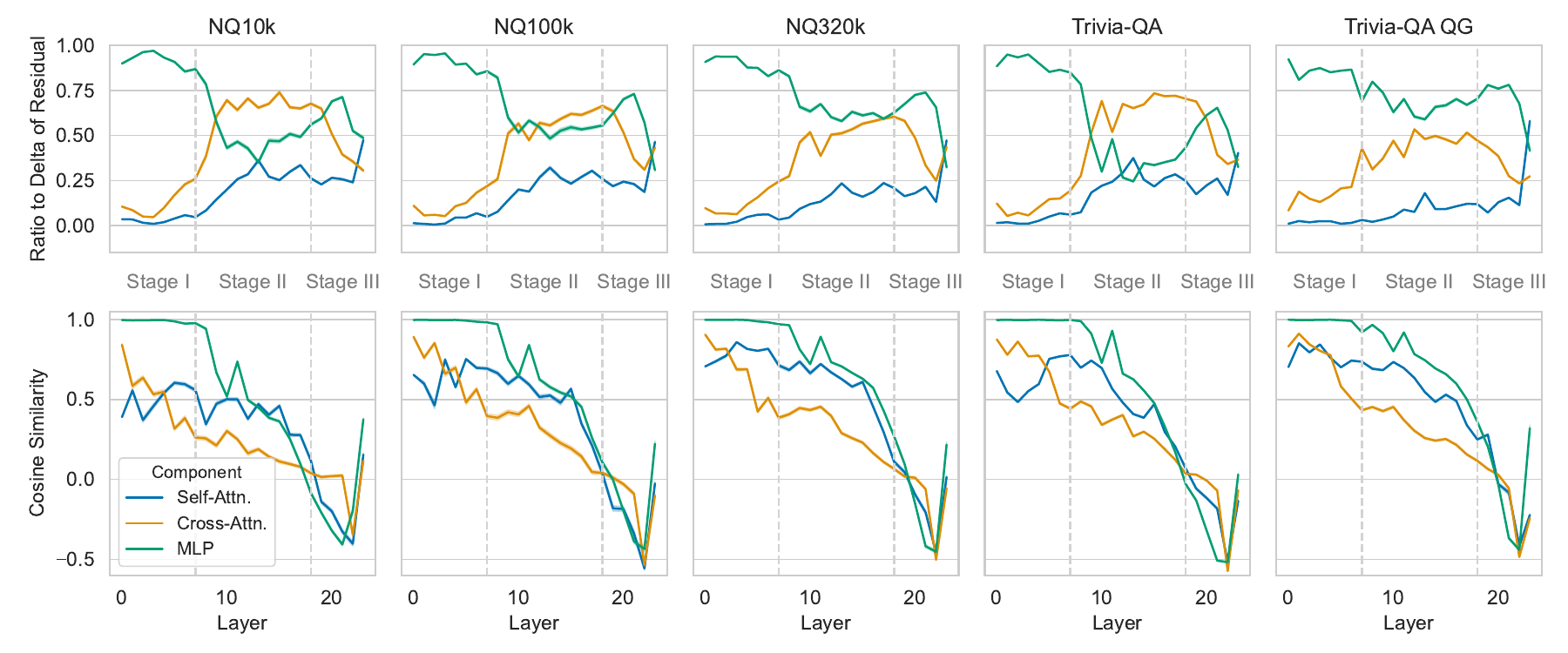}
\caption{Proportion that each component's output contributes to the change in the residual stream (top) and cosine similarity of each  component output with the layer output (bottom), displayed per layer. The models follow a similar trend: high MLP contribution in Stage I and III, cross-attention peaks in Stage II, the cosine similarity of all components is negative in Stage III. }
    \label{fig:cosine-ratio-plots}
\end{figure*}

\textbf{Results}
As shown in Tab.~\ref{tab:results-encoder-swap}, all four models are able to retrieve the removed documents with high performance. For most documents (70 - 80\%), the models are able to rank the removed document at Rank 1 as demonstrated by the high Hits@1 even though the encoder was not optimized for these documents during training. 
Motivated by this high performance, we performed a follow-up experiment and swapped the encoder of the full model with the pre-trained T5-large encoder. The T5 encoder did not receive any GenIR training. When performing retrieval on all queries for which the full model was previously able to place a relevant document at Rank 1, the average rank of the relevant documents drops to $21.66$. For most queries, the relevant document is still placed at Rank 1. Only a few documents are ranked higher than 1000.

\begin{table}[b]
\caption{Results of the Encoder-Swapping Experiment. Models RD I--IV were each trained on 10 documents less (``removed documents'', RD), the full model was trained on the entire corpus.}
\label{tab:results-encoder-swap}

\begin{tabular}{@{}l|rrrrr@{}}
\toprule                                               & Full & RD I   & RD II  & RD III & RD IV  \\ \midrule
MRR & 1    & 0.867 & 0.754 & 0.853 & 0.820 \\
Hits@1 & 10  & 8  & 7  & 8  & 8  \\ \bottomrule
\end{tabular}

\end{table}

Since the models in both experiments are still able to retrieve the documents even though their encoder was not trained to retrieve them, we conclude that the documents are not exclusively encoded in the encoder. We will later see that the information transferred from the encoder to the decoder is mostly similar to non-document-identifiers, which also strengthens the argument that information on document-identifiers are stored in the decoder.
Potentially, the encoder's role lies in semantically encoding the query whose information is then moved to the decoder. Thus, the decoder holds the key for performing retrieval. In the next section, we therefore take a deeper look into the components of the decoder.

\section{Identifying Components Crucial for Retrieval}
\label{sec:components}

The decoder consists of three components: the self-attention, the cross-attention, and the multilayer perceptrons (MLPs). Each model component reads from the residual stream and adds its output back to it. How much each components adds back to the residual stream can be traced by looking at the vector norm of each component output and comparing it to the vector norm  of the residual stream. Since in our atomic doc id scenario, only one token is generated, the residual stream of our decoder consists of only one vector as well. For each component $c$, $c \in \{$Attention, Cross-Attention, MLP~$\}$, we calculate its relative contribution to the residual stream: 
$$
\text{ratio}_c = \frac{\|c_\text{out}\|_2}{\|r_\Delta\|_2}, \text{ with } r_{\Delta} = r_{\text{end}} - r_{\text{begin}},
$$
where $r_\Delta $ is the difference of the residual stream before and after the layer %
and $c_{\text{out}}$ denotes the output of component $c$ that is added to the residual stream. 
This ratio only indicates the magnitude of the contribution of each component, but not whether it adds or subtracts this vector (since the signs of the vector elements could be inverse to the residual stream). Therefore, we also calculate the angle by computing the cosine similarity between the component outputs $c_{\text{out}}$ and the residual stream at the beginning of each layer. 
We aggregate the results of all queries from the validation set for which the model predicted a relevant document at Rank 1 (denoted henceforth as ``correct queries'').\footnote{It is common practice to only evaluate over correctly generated results as we want to investigate here how the model performs retrieval ``correctly''. Using queries for which the retrieval failed would also be of interest as it might point to erroneous behavior the model might have learned, but such analyses are out of scope for this work.}

\paragraph{Results}
Fig.~\ref{fig:cosine-ratio-plots} (top) shows the ratio$_c$ of each model component per layer. 
We see a similar progression for all models: In the beginning, the MLPs dominate, then cross-attention contributes most (or more than before), and in end the MLPs contribute most to the change in the residual. When viewing the cosine similarity (Fig.~\ref{fig:cosine-ratio-plots}, bottom), we see that towards the end all components receive a negative cosine similarity. We suspect that in this part components remove information from the residual stream as their output is directed to the opposite side of the residual stream. Further inspection revealed that the output vectors of these components in this last part indeed have a high negative cosine similarity with components from the first part (lower than -0.9 cosine similarity between the output of each component in the last part and the MLP output of the first part, calculated on the correct queries using NQ10k). 

Based on our observations, we divide the plots into three stages: Stage I (Layers 0--6), where MLPs contribute the most to the residual stream, Stage II (Layers 7--17) where cross-attention receives the highest contribution (positive), and Stage III (Layers 18--23), where MLPs contribute most again, along with attention and cross-attention. In the last stage (excluding the final layer), the contribution vector of each component is negative with respect to the residual stream and the first stage output, suggesting that components in this stage remove information initially contributed by Stage I MLPs. These results indicate which components might be essential in which part of the decoder, as a higher contribution might point to a higher importance. Nevertheless, we do not know whether these components are involved in predicting the output. 
Next, we will therefore causally verify their individual importance.

\subsection{Contribution of Individual Components}
\label{sec:stages-intro}
To investigate whether a model component has a causal influence on the prediction of the model, we apply a variant of activation patching~\cite{vig2020investigating, geiger2021causal}: We replace the output of a certain component at a specified position by zero vector, i.e., removing the component from the computation. Since past research found that this practice might lead to noisy results~\cite{wang2023interpretability}, we also apply mean patching, where we replace a component's output with
the mean output of this component aggregated over correct queries. We measure the effect of our intervention by calculating the rank of the first relevant document after applying the patching, aggregated over all queries where the correct document was originally placed at Rank 1. If the model could not perform retrieval after a component was replaced by a zero vector, we conclude that this component is necessary for the retrieval process. If the model could still perform retrieval after replacing the same component by the mean of the correct queries, this component might be crucial for the retrieval process, but its role is not specific to a certain query, but rather general. In these patching experiments, we use the models trained as described as in Sec~\ref{sec:setup}. We do not re-train models using the indicated components. %
\begin{table*}[]
\caption{Percent of correct queries where the rank of the relevant document is not on Rank 1 after applying patching, we remove/ replace the indicated component output only in the indicated stage. The highest value, i.e., the largest drop in performance per model and stage is indicated bold. MLPs in Stage I can be replaced by mean activations, cross-attention in Stage II and III is crucial for the retrieval process, and MLPs in Stage III are the components adapted most for retrieval.}
\label{tab:patching}

\begin{tabular}{ll rrr rrr rrr}
\toprule
                &        & \multicolumn{3}{c}{NQ10k}      & \multicolumn{3}{c}{NQ320k}     & \multicolumn{3}{c}{Trivia-QA}  \\
                \cmidrule(lr){3-5} \cmidrule(lr){6-8} \cmidrule(lr){9-11}
Component       & Method & Stage I & Stage II & Stage III & Stage I & Stage II & Stage III & Stage I & Stage II & Stage III \\\midrule
Attention       & zero   & 1.52    & 2.27     & 6.57      & 2.67    & 4.46     & 11.23     & 12.58   & 14.78    & 19.97     \\
MLP             & zero   & \textbf{92.86}   & 7.83     & \textbf{96.72}     & \textbf{34.40}   & 30.30    & \textbf{99.28}     & \textbf{29.89}   & 16.94    & \textbf{97.51}     \\
Cross-Attention & zero   & 4.29    & \textbf{21.72}    & 39.65     & 15.86   & 50.62    & 41.89     & 15.29   & \textbf{64.20}    & 60.02     \\\midrule

Attention       & mean   & 0.00    & 2.02     & 6.06      & 1.78    & 5.35     & 11.76     & 14.62   & 16.62    & 20.93     \\
MLP             & mean   & 0.76    & 5.56     & 48.23     & 3.39    & 17.47    & 71.48     & 14.70   & 17.58    & 61.67     \\
Cross-Attention & mean   & 4.55    & 20.45    & 42.93     & 16.04   & \textbf{53.30}    & 50.80     & 18.14   & 54.44    & 71.43     \\\midrule
Attention       & T5     & 5.30    & 5.56     & 7.32      & 10.16   & 11.05    & 13.37     & 5.54    & 6.65     & 9.78      \\
MLP             & T5     & 5.30    & 7.07     & 74.49     & 14.62   & 14.62    & 92.69     & 11.94   & 8.01     & 81.45     \\
Cross-Attention & T5     & 7.58    & 17.93    & 34.09     & 13.19   & 24.96    & 33.16     & 8.91    & 21.31    & 41.38    \\\bottomrule
\end{tabular}
\end{table*}
We patch the three stages we previously identified.
In each stage, we patch each component in all layers of the respective stage.\footnote{Initially, we patched each component individually in each \textit{layer}, but did not observe a drop in performance for the individual components, suggesting that the model could recover its performance, possibly due to redundancy.} %
\paragraph{Results}
Tab.~\ref{tab:patching} summarizes the results for models trained on NQ10k, NQ320k, and Trivia-QA (results for the other models are qualitatively similar). We see approximately the same pattern in all three models. 
In all three stages, the self-attention component can be removed without effecting most of the retrieval performance. In Trivia-QA, replacing self-attention with zero or mean values leads to over 10\% of documents which are not ranked at Rank 1 anymore, suggesting that it has learned a slightly more prominent role for this dataset.
In Stage I, cross-attention does not play an important rule, too, as its performance loss in Stage I is relatively small compared to the other stages.  
In Stages II and III, cross-attention cannot be replaced by zero or mean outputs without a major performance drop. A reason might be its role as the only component that moves query information from the encoder to the decoder. For the models trained on the larger datasets, the highest performance drops caused by patching cross-attention can be seen for Stage II, while for NQ10k the patching in Stage III resulted in higher losses. %
Interestingly, in the first stage, the MLP components can be completely replaced by the mean output of the same components for the correct queries. For all datasets, MLPs contribute substantially to query-specific retrieval in the last stage, resulting in high losses when being replaced or removed. 

Motivated by these results, we perform a subsequent experiment combining the patching of individual stages and components to verify that this partial computation graph of the model can still perform retrieval. 
As we aim to construct the essential combination of components for retrieval, we choose in each stage those components that resulted in the lowest loss in the previous experiment. Concretely, we remove cross-attention from Stage I and attention from all stages, and replace the MLP output in Stages I and II with the mean output over all correct queries. To strengthen the argument that this configuration performs the main computation required by the models for relevance judgment, we evaluate each model on the respective test set.

Tab.~\ref{tab:model-results-patching} shows the results of all models after applying the patch. All models are still able to perform retrieval even though their computation graph contains now one third of the original $72$ active components (3 components---attention, cross-attention, MLPs---per layer $\times$ 24 layers $= 72$). Different models exhibit larger or smaller losses in performance. While for NQ10k the losses are marginal, both Trivia-QA models lose approx.\ 10 points in each of the metrics. %
Nevertheless, all models are still able to perform ranking, demonstrating that these components recover most of the performance of the entire model. We conclude therefore that these model components are essential for the retrieval mechanism.
To summarize, ``active'' components, i.e., components that cannot be removed or replaced by a mean output, are the MLPs in the last stage and cross-attention in the middle and last stages.

\paragraph{Patching in T5}
\label{sec:T5-patching}

The previous experiment showed which components can be removed or replaced by their mean output. However, this does not indicate that these components perform \textit{retrieval-specific} behavior. To investigate this question, we perform patching in the same scenario as before, but this time we replace the output of a component by the output of the same component when the same input is processed by a pre-trained (but not GenIR-trained) T5-large. By doing so, we analyze which mechanisms are learned during pre-training and which are acquired during the GenIR training. This idea is similar to cross-model activation patching from \cite{prakash2024finetuning}, which patched from fine-tuned to pre-trained models, while we patch in the other direction.

The results for patching in T5-large are shown in Tab.~\ref{tab:patching} (bottom). 
Clearly, only a small number of queries require the model's attention components to be trained for GenIR. This is in line with our findings that attention can be removed from the model's computation. Similarly, the MLPs in Stages I and II do not seem to be specifically adapted to the GenIR process, resulting in a relatively small drop of 5--15\% depending on the model. In all three models, the largest drop stems from the MLPs in Stage III, followed by the cross-attention in Stages III and II. 
Cross-attention in Stage II is less adapted to retrieval compared to Stage III, indicating that the Stage II mechanism may have already been acquired during pre-training.

The greatest adaptation to the retrieval task occurs in Stage III, where we see the largest performance change for all three components. We, thus, conclude that Stage III learns information crucial to the retrieval process, potentially specific to the learned corpus.

\begin{table}[t]
\caption{Results on the respective test set after patching. Stage~I consists of MLP mean outputs, Stage~II of MLP mean outputs and cross-attention, and Stage~III contains both cross-attention and MLPs. Percentages denote results relative to the model without interventions. On all datasets, ranking can still be performed efficiently even though all models suffer small losses.}
\label{tab:model-results-patching}
\begin{tabular}{@{}l rr rr rr@{}}
\toprule
             & \multicolumn{2}{c}{Hits@1} & \multicolumn{2}{c}{R@5} & \multicolumn{2}{c}{Hits@10} \\ 
             \cmidrule(lr){2-3} \cmidrule(lr){4-5} \cmidrule(lr){6-7}
             & Abs.         & Rel.        & Abs.        & Rel.      & Abs.          & Rel.        \\\midrule
NQ10k        & 38.0         & 94\%        & 56.90       & 98\%      & 62.0          & 97\%        \\
NQ100k       & 17.60        & 78\%        & 35.10       & 82\%      & 41.20         & 82\%        \\
NQ320k       & 16.40        & 82\%        & 33.60       & 84\%      & 42.01         & 88\%        \\
Trivia-QA    & 49.67        & 82\%        & 55.26       & 86\%      & 74.95         & 91\%        \\
Trivia-QA QG & 38.94        & 78\%        & 41.60       & 80\%      & 57.97         & 87\%        \\ \bottomrule
\end{tabular}
\end{table}

\begin{figure*}[t]
    \centering
\includegraphics[width=\textwidth]{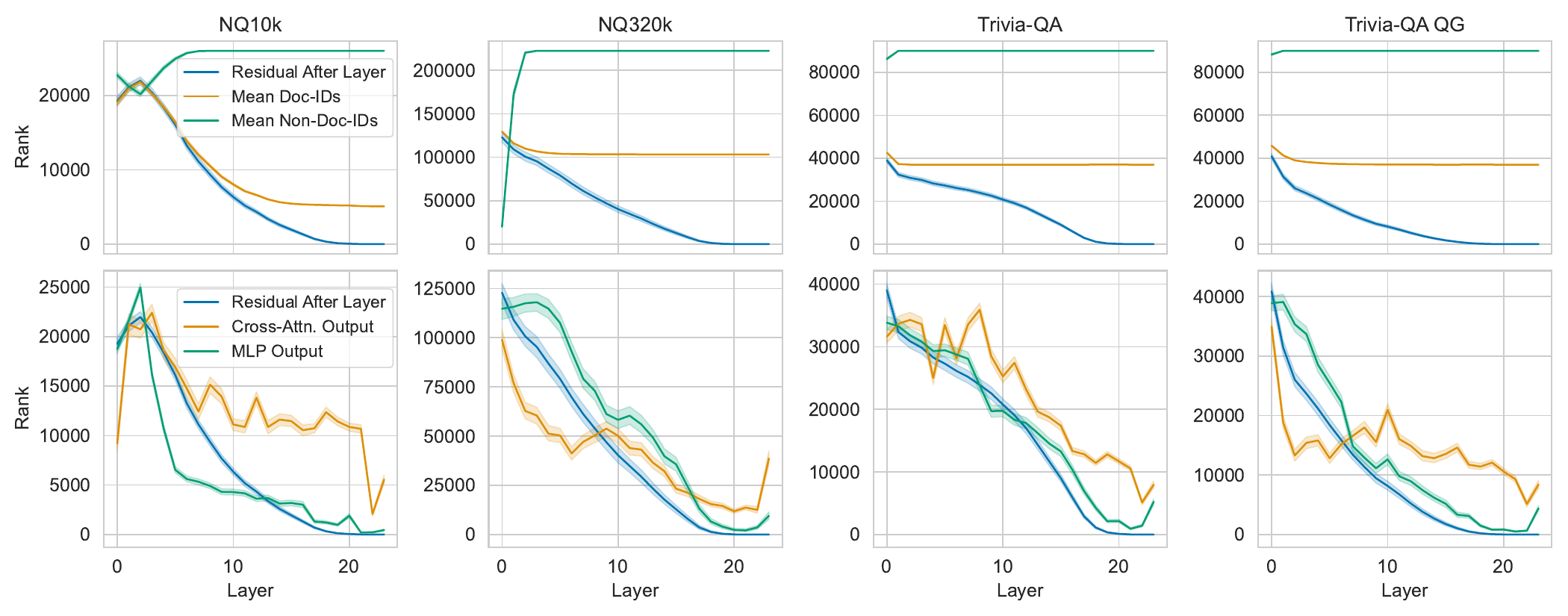}
    \caption{Rank of the correct document after each layer, average rank of all document identifier tokens and non-document-identifier tokens after applying logitlens to the output of the layer (top), and rank of the correct document after applying logitlens to the indicated model component (bottom), displayed per layer. All models follow a similar trend (including NQ100k, omitted for space concerns): The models separate document identifiers and non-document-identifiers early, and gradually improve the rank of the relevant document. The cross-attention output does not seem to follow this gradual progression.}
    \label{fig:rank-plots}
\end{figure*}

\subsection{Rank Development}
\label{sec:rank-developement}

Recall that the objective of a GenIR model is to predict the relevant document identifier, which means it is optimized to place the relevant document identifier at a low rank in its output. 
In this section, we examine the changes in the rank of the relevant document identifier caused by different model components. This analysis provides a first intuition on the role of the MLPs and cross-attention.

We adopt the \textit{logit lens}~\cite{nostalgebraist2020logit}, a technique that projects intermediate hidden states from a model to the space of output vocabulary tokens. This technique has been used extensively to study how decoder language models build representations when processing a given input \cite{geva-etal-2022-lm,geva2021transformer,dar-etal-2023-analyzing,katz2023visit,katz-etal-2024-backward}. 
The key intuition is that due to the residual stream view, hidden states at different layers can be projected to the output vocabulary with the unbembedding matrix: 
$$
\textit{logit-lens}(r) = W_U \cdot \text{scale}(\text{ReLU}(r)),$$
where scale($\cdot$) scales its input by $\sqrt{d_{\text{model}}}$ as done before the unembedding layer in T5 models. logit-lens$_i(r)$ contains the logits of the $i$th token in the vocabulary when applied to vector $r$. The higher the logit value, the lower the rank of the corresponding token, such that the token at Rank 1 has the highest logit value. We use logit-lens to determine the ranks of document id tokens, of non-document-id tokens, and of relevant document ids, when applying it to the outputs of cross-attention, MLPs, and the residual stream in total after each layer. We average the ranks over all correct queries. 
In this experiment, we use all model components and do not remove or replace model outputs as in the previous section.

\paragraph{Results}
Fig.~\ref{fig:rank-plots} displays the development of the rank for the relevant documents for each layer (top) and the outputs of cross-attention and MLPs per layer (bottom).
In early layers, the relevant documents moves to the lower half of all ranks together with the average rank of all document ids. It gradually decreases until it arrives at Rank~1 in the last layer. A similar progression can be seen for the MLP output. But for the cross-attention output, the rank of the relevant document does not gradually change. Instead, there is no clear trend visible.

One hypothesis is that only the MLPs contribute directly to the ranking of the documents by modifying the residual stream. 
We therefore calculate the rank of each document when only the output of MLPs contributes to the logits. In this experiment, for each query, we perform a pass through the model, so each model component uses its ``normal'' input to calculate its output without any intervention. Only the calculation of the logits is changed from using the residual stream with the output of all components (attention, cross-attention, and MLPs) to using only the residual steam consisting of the MLP outputs.
As before, the experiment is performed on the correct queries only. Here, we only use NQ10k.
When only using the output of MLPs to calculate the logits, the performance drops slightly: 33\% of the relevant documents are not placed at Rank 1 anymore, but instead between Rank 1 and 10, indicating that ranking can still be performed well. Performing the same experiment with only using the output of the cross-attention components reveals that ranking using this information only is not possible. For all correct queries, the relevant document is ranked above 100. 

The results of both analyses indicate that cross-attention does not directly influence the rank of document identifiers, but instead passes information to the MLPs, which then ``write'' the relevant information to the residual stream. This communication between cross-attention and MLPs is analyzed in detail in the next section.

\subsection{Communication between Components}
\label{sec:mlp-cr-communication}
We saw in the previous sections that we can divide the pass through the decoder into three stages. In Stage II, cross-attention contributed more than other components, while in Stage III, the MLPs seem to be the most important components. In addition, the MLP components, as demonstrated in the previous section, mainly affect the output logit and, thereby, the output ranking. Therefore, we suspect that communication between these two component types is a crucial part of the retrieval process.
To investigate this communication between cross-attention and MLPs, we want to gather from which components the MLPs in Stage III and the cross-attention in later stages read. 
The intuition behind the following analyses is that the input to MLPs and cross-attention is the residual stream, which is the sum of all component outputs before the component which we investigate. Therefore, we can compute the contribution of each component to the MLPs/cross-attention output to determine the most influential components.
\paragraph{Contribution to MLP Output}
Recall from Sec.~\ref{sec:background} that only positively activated neurons contribute to the MLP output. We therefore investigate how these neurons get activated. The $i$th neuron is activated when $\smash{\text{FF}^\text{proj}_i \cdot r > 0}$. The MLP's input is the residual stream $r$, which is the sum of all previous components. For each activated neuron, we iterate through all previous components and calculate the contribution $t_\text{MLP}$ as the dot product between $\smash{\text{FF}^\text{proj}_i}$ and each component's output $c_\text{out}$: $\smash{t_\text{MLP}(c_\text{out}) = \text{FF}^\text{proj}_i \cdot c_\text{out}}$. We average over all activated neurons and all correct queries.

\paragraph{Contribution to Cross-Attention Output}
In a similar way, we calculate the contribution of all components that lead to the contribution of a cross-attention head. 
The composition of the output of one cross-attention head is determined by the similarity-scores $(s_1, ..., s_N)$. Let $\hat{i}$ denote the index of the highest score:
$\hat{i} = \arg\max_i(s_i)$. 
Because the highest similarity score $s_{\hat{i}}$ was computed by $q^T\cdot k_{\hat{i}} = (W_Q \cdot r)^T\cdot k_{\hat{i}}$, 
we iterate though all previous component outputs $c_{\text{out}}$ (which are the ones the residual stream is composed of) and calculate the contribution score $t_{\text{Cr-Attn}}$ as: 
$t_{\text{Cr-Attn}}(c_{\text{out}}) = W_Q \cdot c_\text{out} \cdot k_{\hat{i}}$. We average over all heads  and all correct queries.

\paragraph{Results}

\begin{figure}
    \centering
    \includegraphics[width=0.99\linewidth]{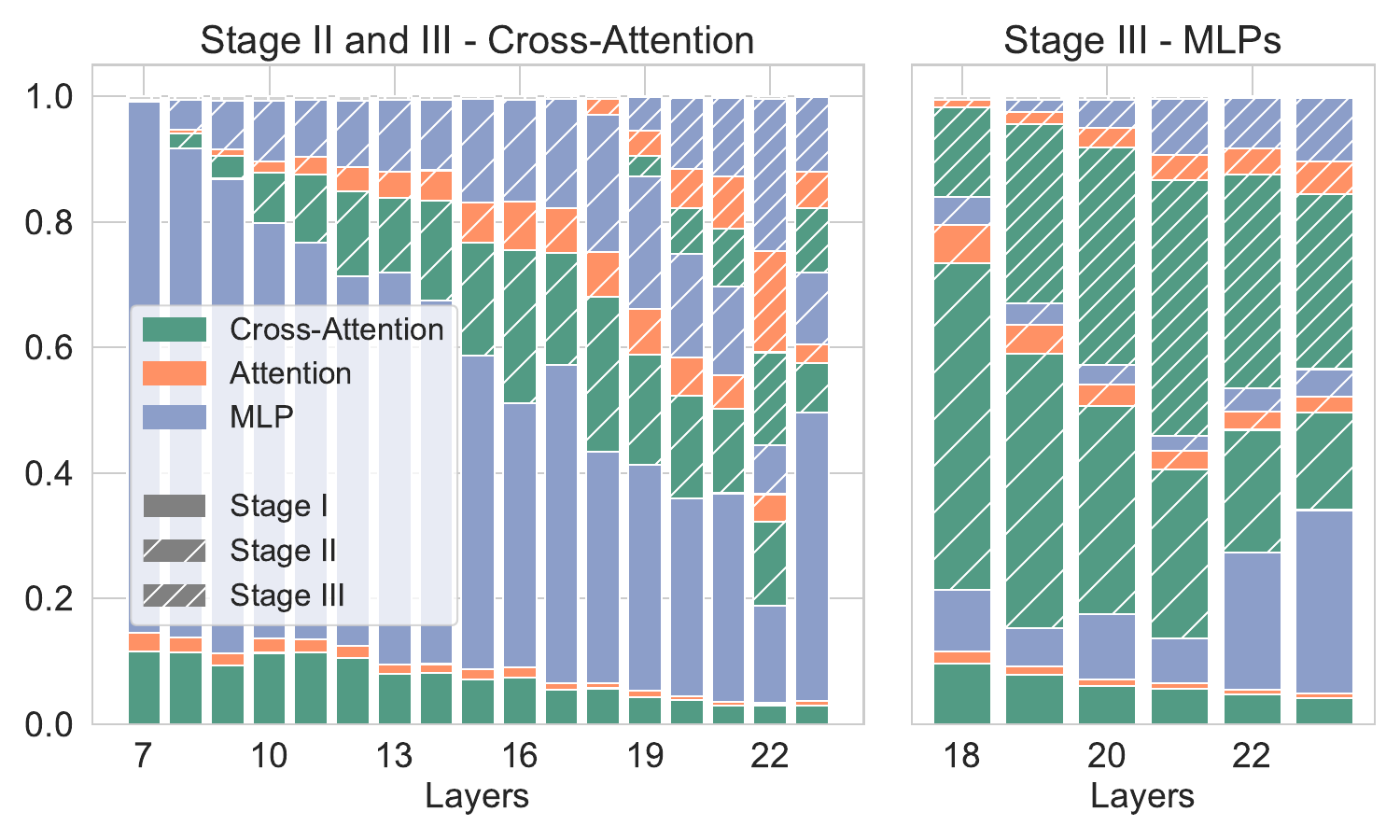}
    \caption{Components per stage that trigger cross-attention in Stage II and III (left) and activate MLPs in Stage III (right) of NQ10k. Stage III MLPs gets mostly activated from cross-attention in Stage II and III, while cross-attention in Stage II gets mostly activated by Stage I MLPs.}
    \label{fig:component-readings}
\end{figure}

For each correct query, we compute the active neurons in NQ10k and determine the model components with the highest contribution to the activation. Fig.~\ref{fig:component-readings} displays the proportions of the dot product scores of each component to the cross-attention of Stage II and III, $t_{\text{Cr-Attn}}$,  and the active neurons of the MLPs of Stage III, $t_{\text{MLP}}$. We %
aggregate the components by type and stage. Each bar in the diagram shows from which previous components the cross-attention (left) or MLP (right) in this layer reads.  
The results show that cross-attention from Stage II and later also from Stage III plays a major role in shaping the output of the MLPs, as their output is mostly determined by the information which these components output (proportion of green colors in each bar in the right plot). 
The output of the cross-attention components in Stage II and III get mostly determined by the MLPs from Stage I (proportion of solid blue color in the left plot). In later layers, other components from Stage II and III also contribute, but still less than the Stage I MLPs. Layer 22 is an outlier as Stage III MLPs contribute the largest share to trigger the cross-attention heads in this layer. 

In addition, we repeat this analysis, but for individual cross-attention heads. We find that for NQ10k the same heads (66\% of all cross-attention heads) are within the top 5 components with the highest contribution for all correct queries. Overall, all cross-attention heads except for the last layer ones appeared in the top 5 for at least  one query. This finding indicates that all cross-attention heads participate in computing the output (through MLPs).

Overall, these analyses show that MLPs in Stage~III, which mainly contribute to the final result, get activated by the cross-attention output (in Stage~II and later Stage~III). In contrast, the cross-attention output is mostly determined by MLPs from Stage I. 
These results suggest that Stage I serves as a \textit{priming} or preparation phase to contribute information to the residual stream, which later triggers cross-attention. Cross-attention then transfers information from the encoder to the residual stream. This information activates neurons in Stage III. Nevertheless, it remains open what cross-attention writes and the neurons read. We take a look at this aspect next. %

\begin{table}[]
\caption{Percentage of Document-Identifier tokens within the Top 10, 100 and 1,000 tokens with highest logits in the logit-lens projection of the Cross-Attention Components. Average over all layers in Stage II, all heads, and all correct queries.}
\label{tab:cross-attn-output-proportion}
\begin{tabular}{@{}lrrr@{}}
\toprule
Dataset      & Top 10 & Top 100 & Top 1,000\\ \midrule
 NQ10k        &    0.00    &    0.10    & 2.30 \\
 NQ100k       &   0.00     &    0.01     & 0.50\\
 NQ320k       &   0.00     &     0.04    & 0.57\\
 Trivia-QA    &    0.00    &     0.08    & 8.58\\
 Trivia-QA QG &    0.02    &    0.16     &2.68\\\bottomrule
\end{tabular}
\end{table}

\subsection{Cross-Attention and MLPs communicate in the word-token space}
\label{sec:communication-in-word-space}
To better understand how MLPs and Cross-Attention communicate, we apply logit-lens (see Sec.~\ref{sec:rank-developement}) to the output of the cross-attention components. This way, we essentially identify which tokens the cross-attention output is most similar to. We sort all tokens in the vocabulary of the model by their logit value, and divide them into document identifier tokens and non-document-identifier tokens. We then compute the proportion of non-document-identifiers within the top 10, 100, and 1,000 tokens with highest logits. The results (Tab.~\ref{tab:cross-attn-output-proportion}) show that cross-attention heads mostly output vectors that are more similar to non-document-identifier tokens. Only in rare cases do document-identifiers appear in the top 1,000. Since we saw that MLPs are activated by these outputs, these results suggest that cross-attention and MLPs communicate in the space of non-document-identifier tokens.

\begin{table*}[]
\caption{Examples for top 5 words whose logits were promoted by a given head when the model received the query as its input, computed on NQ10k.}
\label{tab:example-words}
\begin{tabular}{@{}lll@{}}
\toprule
Query                                     & Head               & Top 5 Words                                        \\ \midrule
\multirow{2}{*}{who wrote the harry potter books}  & Layer 14 - Head 2  & about, written, about, tailored, privire           \\     
& Layer 16 - Head 1  & books, ouvrage, books, authors, book               \\  \midrule
\multirow{2}{*}{who won the football championship in 2002} &  Layer 16 - Head 1  & year, YEAR, Year, year, jahr                      \\ 
                                        & Layer 16 - Head 13 & football, Football, fotbal, soccer, NFL            \\\midrule
\multirow{2}{*}{will there be a sequel to baytown outlaws} & Layer 12 - Head 8  & erneut, successor, similarly, repris, continuation \\
                                          & Layer 16 - Head 1  & town, towns, city, Town, village                   \\ \bottomrule
\end{tabular}

\end{table*}

In addition, in Tab.~\ref{tab:example-words} we plot the top 5 tokens for three queries along the middle layers after projecting the cross-attention head outputs in these layers using logit-lens. Evidently, these tokens are semantically similar to the input query. Among them are also tokens in other languages like German or French that share similar semantics. It seems that the model has learned to perform query expansion similar to other neural retrieval models~\cite{ram2023you}.

This might indicate that the communication between cross-attention and MLPs happens in the space of word-tokens, and that cross-attention learns no information on the document ids or propagates such information. 
We saw in Sec.~\ref{sec:T5-patching} that the model acquired several mechanisms that are part of the retrieval process already during pre-training. %
Therefore, the communication in the word space could also be a left-over from the next word prediction task, which the model was originally trained on during pre-training.

\section{The Retrieval Process in GenIR}
\label{sec:retrieval-walkthrough}

This section summarizes our findings and provides a walkthrough of the retrieval process we discovered in our GenIR models. A high-level overview of the retrieval process is displayed in Fig.~\ref{fig:overview-figure}.

\textit{Encoder.}
The model begins by embedding the query at the beginning of the encoder. The embedded query vectors are passed through the encoder, which applies MLPs and self-attention to contextualize the input tokens. Our results show that the encoder is not required to encode information on the documents directly as it can be replaced by an encoder that does not contain document specific information (see Sec.~\ref{sec:encoder-swap}). 

\textit{Priming Stage.}
The decoder receives as its input a generic start token, which also gets embedded in the same way as in the encoder. In the first stage  of the model (layers 0--6), no query specific information is required (Sec.~\ref{sec:stages-intro}). Using the MLP components, the model moves document id tokens to lower ranks and non-document-id tokens to higher ranks~(Sec.~\ref{sec:rank-developement}) while adding information used by subsequent layers to trigger cross-attention.

\textit{Bridging Stage.} In the second stage (layers 7--17), cross-attention moves information from the encoder to the decoder. The cross-attention heads output information on the input query to the residual stream, which resembles a form of query expansion. 
This information is then used to activate neurons in later layers (Sec.~\ref{sec:mlp-cr-communication}). Our experiments indicate that this communication takes places in the space of word-tokens (Sec.~\ref{sec:communication-in-word-space}). 

\textit{Interaction Stage.}
In Stage III (layers 18--23), cross-attention continues to output query information to the residual stream. 
At the same time, the MLP neurons are activated and output information that promotes document identifier tokens (Sec.~\ref{sec:rank-developement}). 
In the last layer, only the MLPs are required. They do not read from the last layer Cross-Attention component (Sec.~\ref{sec:mlp-cr-communication}). In this layer, all non-document-id tokens are moved to lower ranks, such that only document id tokens will be predicted by the model (Sec.~\ref{sec:rank-developement}). This stage is the one that received the greatest adaption to the retrieval task during the GenIR training. This fact implies that corpus specific information is most probable be stored in components in this stage. Therefore, the query interacts only in this stage with document information. Finally, in the last layer, the residual stream is multiplied by the unembedding matrix, resulting in logits for each token. Relevant document identifier tokens should receive the highest logits for retrieval to be successful.

Even though our results suggest that the same retrieval process is present in all models we examined, differences were observable as well. In models trained on larger datasets, MLPs in the first and second stage seem similarly important, indicating that the priming stage might be longer than seven layers as for NQ10k.

\section{Conclusion}
\label{sec:conclusion}

In this work, we reverse-engineered the retrieval process in atomic GenIR models using mechanistic interpretability methods. %
First, we performed retrieval using an encoder trained on fewer or no documents. This experiment demonstrated that the retrieval process within GenIR models does not rely on a retrieval-tuned encoder. 
Our main contribution is uncovering and analyzing the mechanism behind the retrieval process. We showed that the pass through the decoder can be divided into three stages: (I) the priming stage, which prepares the residual stream for subsequent components; (II) the bridging stage, where cross-attention is primarily active to transfer query information from the encoder to the decoder; and (III) the interaction stage, where predominantly MLPs are active to promote relevant documents. Furthermore, our patching experiments revealed that GenIR models can still rank documents effectively even with only one-third of their components.

These findings indicate that only a small part of the models needs adaptation for effective GenIR. We provided evidence that several mechanisms within the retrieval process, such as the priming and the bridging stages, are learned as part of the pre-training. This suggests that GenIR fine-tuning can be optimized by adjusting only the necessary components for faster training, or forcing to adapt all components for better retrieval effectiveness.
Additionally, we found that several components are agnostic to the query since they can be replaced by a generic activation of all queries. Other components such as attention can be removed completely, eliminating the need to re-compute them during inference. This suggests potential optimizations for inference efficiency.

Our experiments revealed the retrieval process in atomic GenIR models, supported by evidence from two datasets in different variations. However, we have not uncovered \textit{how} models learn mechanisms from the data, so we cannot guarantee that every GenIR model will operate in this manner. It seems that part of the process we uncovered was already acquired by the model during pre-training. The retrieval process reuses these existing mechanisms and adapts necessary parts, particularly in the MLPs of Stage III. Thus, we suspect that different pre-training schemes may lead to distinct retrieval mechanisms that models learn. Therefore, future work should focus on analyzing the impact of training data and differently pre-trained models on the retrieval mechanism. 
Overall, this work has taken a first step toward elucidating the inner workings of GenIR models. To facilitate further research we release our models and code.

\begin{acks}
The authors are grateful to Maik Fröbe and Hadas Orgad for providing feedback on this work. 
A.R. was funded through the Azrieli international postdoctoral fellowship and the Ali Kaufman postdoctoral fellowship.

This research was supported by an Azrieli Foundation Early Career Faculty Fellowship, and an AI Alignment grant from Open Philanthropy.
This research was partly funded by the European Union (ERC, Control-LM, 101165402). Views and opinions expressed are however those of the author(s) only and do not necessarily reflect those of the European Union or the European Research Council Executive Agency. Neither the European Union nor the granting authority can be held responsible for them.
\end{acks}

\bibliographystyle{ACM-Reference-Format}
\bibliography{references}

\end{document}